\documentclass[reqno]{amsart}
\usepackage{graphicx}
\usepackage{fullpage}

\theoremstyle{definition}

\newcommand{\green}{f_{_{\rm G}}}
\def\ngreen{n_{_{\rm G}}}
\def\ugreen{U_{_{\rm G}}}

\def\sigpar{\sigma_\|}
\def\sigperp{\sigma_\perp}

\def\ymin{y_{\rm min}}
\def\ymax{y_{\rm max}}

\theoremstyle{remark}

\begin{document}

\title[Hypergeometric Green's function]
{Exact solution for the hypergeometric Green's function
describing spectral formation in X-ray pulsars}

\author[Peter A. Becker]{Peter A. Becker\\
Center for Earth Observing and Space Research,\\
School of Computational Sciences,\\
George Mason University,\\
Fairfax, VA 22030-4444, USA}

\email{pbecker@gmu.edu}


\date{Submitted December 11, 2004.}


\vskip2.0truein
\centerline{accepted by the Journal of
Mathematical Physics}

\begin{abstract}
An eigenfunction expansion method involving hypergeometric functions is
used to solve the partial differential equation governing the transport
of radiation in an X-ray pulsar accretion column containing a radiative
shock. The procedure yields the exact solution for the Green's function,
which describes the scattering of monochromatic radiation injected into
the column from a source located near the surface of the star.
Collisions between the injected photons and the infalling electrons
cause the radiation to gain energy as it diffuses through the gas and
gradually escapes by passing through the walls of the column. The
presence of the shock enhances the energization of the radiation and
creates a power-law spectrum at high energies, which is typical for a
Fermi process. The analytical solution for the Green's function provides
important physical insight into the spectral formation process in X-ray
pulsars, and it also has direct relevance for the interpretation of
spectral data for these sources. Additional interesting mathematical
aspects of the problem include the establishment of a closed-form
expression for the quadratic normalization integrals of the orthogonal
eigenfunctions, and the derivation of a new summation formula involving
products of hypergeometric functions. By taking various limits of the
general expressions, we also develop new linear and bilinear generating
functions for the Jacobi polynomials.
\end{abstract}

\maketitle

\section*{\bf I. INTRODUCTION}

In this article, methods of classical analysis are employed to obtain
the exact solution for the Green's function describing the spectrum of
radiation emitted by an X-ray pulsar. Beyond the direct physical
relevance of the Green's function, the method of solution also yields
several additional results of mathematical interest, including a new
summation formula involving products of two hypergeometric functions, as
well as new linear and bilinear generating functions for the Jacobi
polynomials. We also obtain an exact expression for the quadratic
normalization integrals of the orthogonal hypergeometric eigenfunctions.
Before proceeding with the main derivation, some physical background is
called for. The radiation produced in bright X-ray pulsars is powered by
the gravitational accretion (inflow) of ionized gas that is channeled
onto the poles of a rotating neutron star by the strong magnetic field.
In these sources, the radiation pressure greatly exceeds the gas
pressure, and therefore the pressure of the photons governs the
dynamical structure of the accretion flow. It follows that the gas must
pass through a radiation-dominated shock on its way to the stellar
surface, and the kinetic energy of the gas is carried away by the
high-energy radiation that escapes from the column.$^{\ref{ref1}}$ The
strong gradient of the radiation pressure decelerates the material to
rest at the surface of the star, and the compression of the infalling
gas drives its temperatures up to a few million Kelvins. The gas
therefore radiates X-rays, which appear to pulsate due to the star's
spin. However, the observed X-ray spectrum is nonthermal, indicating
that nonequilibrium processes are playing an important role in the
formation of the radiation distribution.

The nonthermal shape of the spectrum is primarily due to the flow
compression, which causes Fermi energization of the photons as they
collide with infalling electrons in the column, until the radiation
escapes from the column into space. Our primary goal in this article is
to obtain an exact solution for the Green's function describing the
upscattering of soft, monoenergetic photons injected by a source located
in the base of the accretion column, near the surface of the star. The
Green's function contains a complete representation of the fundamental
physics governing the propagation of the photons in the physical and
energy spaces. Since the transport equation governing the radiation
distribution is linear, we can compute the solution associated with an
arbitrary source distribution via convolution. Hence the Green's
function provides the most direct means for exploring the relationship
between the physics occurring in the accretion shock and the production
of the observed nonthermal X-radiation.

\section*{\bf II. FUNDAMENTAL EQUATIONS}

We assume that the accretion column is cylindrical, and we define $x$ as
the spatial coordinate measured along the column axis. The gas flows
through the column onto the stellar surface with velocity $v$. We define
the Green's function, $\green(x_0,x,\epsilon_0,\epsilon)$, as the
radiation distribution at location $x$ and energy $\epsilon$ resulting
from the injection of $\dot N_0$ photons per second with energy
$\epsilon_0$ from a monochromatic source at location $x_0$ inside the
column. In a steady-state situation, $\green$ satisfies the transport
equation$^{\ref{ref2},\,\ref{ref3}}$
\begin{equation}
v{\partial\green\over\partial x}={dv\over d x}\,{\epsilon\over 3}\,
{\partial\green\over\partial\epsilon} + {\partial\over\partial x}
\left({c\over 3 n_e \sigma_\|}\,{\partial \green\over\partial x}\right)
+ {\dot N_0 \, \delta(\epsilon-\epsilon_0) \, \delta(x-x_0)\over \pi r_0^2
\epsilon_0^2}
- {\green \over t_{\rm esc}}
- \beta \, v_0 \, \delta(x-x_0) \, \green
\ ,
\label{eq2.1}
\end{equation}
where $n_e$ is the electron number density, $\sigpar$ is the electron
scattering cross section for photons propagating parallel to the
$x$-axis, $r_0$ is the radius of the column, $v_0$ is the flow speed at
the source location, $c$ is the speed of light, and $t_{\rm esc}$ is the
mean time photons spend in the column before escaping through the walls
into space. The total radiation number and energy densities associated
with the distribution function $\green$ are, respectively,
\begin{equation}
\ngreen(x) \equiv \int_0^\infty \epsilon^2 \,
\green \, d\epsilon \ , \ \ \ \ \ 
\ugreen(x) \equiv \int_0^\infty \epsilon^3 \,
\green \, d\epsilon
\ .
\label{eq2.2}
\end{equation}
The terms in (\ref{eq2.1}) represent, from left to right, the comoving
(convective) time derivative, first-order Fermi energization (``bulk
Comptonization'') of the radiation in the converging flow, spatial
diffusion of the photons parallel to the column axis, the monochromatic
photon source, escape of radiation from the column, and the possible
absorption of radiation at the source location, respectively. In
physical terms, the first-order Fermi energization corresponds to the
$PdV$ work done on the radiation by the compression of the background
plasma as it accretes onto the stellar surface.$^{\ref{ref3}}$ The
dimensionless constant $\beta$ expresses the strength of the absorption
(if any) occurring at the source location, and the mean escape time is
given by
\begin{equation}
t_{\rm esc} = {r_0^2 \, n_e \, \sigperp \over c}
\ ,
\label{eq2.3}
\end{equation}
where $\sigperp$ is the electron scattering cross section for photons
propagating perpendicular to the column axis. In general, $\sigpar \ne
\sigperp$ due to the influence of the strong magnetic field, which is
directed parallel to the axis of the column. Absorption at the source
location is expected if the photons are produced in a blackbody
``mound'' of dense gas near the base of the accretion
column,$^{\ref{ref1}}$ because a perfect blackbody acts as both a source
and a sink of radiation.$^{\ref{ref4}}$

The flux of electrons flowing down the column is denoted by $J \equiv
n_e v$. In our cylindrical, steady-state problem, $J$ maintains a
constant value. Becker$^{\ref{ref5}}$ demonstrated that in order for the
inflowing matter to come to rest at the stellar surface as required, the
parameters $r_0$, $J$, $\sigpar$, and $\sigperp$ must satisfy the
dynamical constraint
\begin{equation}
r_0^2 \, J^2 \, \sigperp \, \sigpar = {3 \over 4} \, c^2
\ .
\label{eq2.4}
\end{equation}
In general, radiation-dominated shocks are continuous velocity
transitions, with an overall thickness of a few Thomson scattering
lengths, unlike standard (discontinuous) gas-mediated
shocks.$^{\ref{ref6}}$ The exact solution for the inflow velocity $v$ as
a function of the spatial coordinate $x$ is given
by$^{\ref{ref5},\,\ref{ref7}}$
\begin{equation}
{v(x) \over v_c} = {7 \over 4} \left[1 - \left(7 \over 3\right)
^{-1+x/x_{\rm st}}\right]
\ ,
\label{eq2.5}
\end{equation}
where $v_c$ is the flow velocity at the sonic point, which is
related to the stellar mass $M_*$, the stellar radius $R_*$,
and the gravitational constant $G$ via$^{\ref{ref5}}$
\begin{equation}
v_c = {4 \over 7} \left(2 \, G M_* \over R_*\right)^{1/2}
\ .
\label{eq2.6}
\end{equation}
The quantity $x_{\rm st}$ appearing in (\ref{eq2.5}) is the distance
between the sonic point and the stellar surface, which can be evaluated
using Eq.~(4.16) from Ref.~4 to obtain
\begin{equation}
x_{\rm st} = {r_0 \over 2 \sqrt{3}} \left(\sigperp \over \sigpar
\right)^{1/2} \ln\left(7 \over 3\right)
\ .
\label{eq2.7}
\end{equation}
According to (\ref{eq2.5}), the flow does come to rest at the surface of
the star as required, since $v(x_{\rm st})=0$. Furthermore, the
constancy of the electron flux $J$ in our cylindrical, steady-state
problem implies that the electron number density $n_e$ is a function of
$x$ because $v$ varies with the height inside the column [see
Eq.(\ref{eq2.5})].

Further simplification is possible if we work in terms of the new
spatial variable $y$, defined by
\begin{equation}
y(x) \equiv \left(7 \over 3\right)^{-1 + x/x_{\rm st}}
\ .
\label{eq2.8}
\end{equation}
Note that $y \to 0$ in the far upstream region ($x \to -\infty$), and $y
\to 1$ at the surface of the star ($x \to x_{\rm st}$). Based on
(\ref{eq2.5}) and (\ref{eq2.8}), we find that the variation of the
velocity $v$ as a function of the new variable $y$ is given by the
simple expression
\begin{equation}
{v(y) \over v_c} = {7 \over 4} \, (1-y)
\ .
\label{eq2.9}
\end{equation}
By combining (\ref{eq2.3}), (\ref{eq2.4}), and (\ref{eq2.9})
with the derivative relation
\begin{equation}
{dx \over dy} = {r_0 \over 2 \sqrt{3}} \, \left(\sigperp \over
\sigpar\right)^{1/2} y^{-1}
\ ,
\label{eq2.10}
\end{equation}
we can transform the transport equation~(\ref{eq2.1}) for $\green$
from $x$ to $y$ to obtain
\begin{eqnarray}
y \, (1-y) \, {\partial^2 \green \over \partial y^2}
&+& \left({1 - 5 \, y \over 4}\right) {\partial \green \over \partial y}
- {\epsilon \over 4} \, {\partial \green \over \partial \epsilon}
+ \left(y - 1 \over 4 \, y \right) \green
\nonumber
\\
&=& {3 \, \beta \, v_0 \, \delta(y-y_0) \, \green \over 7 \, v_c}
- {3 \, \dot N_0 \, \delta(\epsilon-\epsilon_0) \, \delta(y-y_0)
\over 7 \, \pi \, r_0^2 \, \epsilon_0^2 \, v_c}
\ ,
\label{eq2.11}
\end{eqnarray}
where $y_0 \equiv y(x_0)$ denotes the value of $y$ at the source
location. According to (\ref{eq2.9}), the flow velocity at the
source, $v_0$, is related to $v_c$ and $y_0$ by
\begin{equation}
{v_0 \over v_c} = {7 \over 4} \, (1 - y_0)
\ .
\label{eq2.12}
\end{equation}
Note that we can write the Green's function as either $\green(x_0,x,
\epsilon_0,\epsilon)$ or $\green(y_0,y,\epsilon_0,\epsilon)$ since
the parameters $(x,x_0)$ and $(y,y_0)$ are interchangeable via
(\ref{eq2.8}).

\section*{\bf III. SOLUTION FOR THE GREEN'S FUNCTION}

The physical model considered here includes Fermi energization, which
tends to boost the energy of the injected photons as they collide with
high-energy electrons streaming down through the accretion column
towards the surface of the neutron star. Moreover, since no process that
can lower the photon energy is included in the model, all of the photons
injected from a source of monochromatic radiation with energy $\epsilon
= \epsilon_0$ must at later times have energy $\epsilon > \epsilon_0$.
It follows that $\green = 0$ for $\epsilon < \epsilon_0$. When $\epsilon
> \epsilon_0$, (\ref{eq2.11}) is separable in energy and space using the
functions
\begin{equation}
f_\lambda(\epsilon,y) = \epsilon^{-\lambda} \, g(\lambda,y) \ ,
\label{eq3.1}
\end{equation}
where $\lambda$ is the separation constant, and the spatial function
$g$ satisfies the differential equation
\begin{equation}
y \, (1-y) \, {d^2 g \over dy^2}
+ \left({1 - 5 \, y \over 4}\right) {d g \over dy}
+ \left({\lambda \, y + y - 1 \over 4 \, y}\right) g
= {3 \, \beta \, v_0 \, \delta(y-y_0) \over 7 \, v_c} \, g
\ .
\label{eq3.2}
\end{equation}
In order to avoid an infinite spatial diffusion flux at $y=y_0$, the
function $g$ must be continuous there, and consequently we obtain the
condition
\begin{equation}
\Delta\left[g(\lambda,y)\right]\Bigg|_{y=y_0}
\equiv \ \lim_{\varepsilon \to 0} g(\lambda,y_0+\varepsilon)
- g(\lambda,y_0-\varepsilon) \ = \ 0
\ .
\label{eq3.3}
\end{equation}
We can also derive a jump condition for the derivative $dg/dy$ at the
source location by integrating (\ref{eq3.2}) with respect to $y$ in a
small region around $y=y_0$. The result obtained is
\begin{equation}
\Delta\left[{d g \over d y}\right]\Bigg|_{y=y_0}
= {3 \, \beta \over 4 \, y_0} \ g(\lambda,y_0)
\ ,
\label{eq3.4}
\end{equation}
where we have used (\ref{eq2.12}) to substitute for $v_0$.

The homogeneous version of (\ref{eq3.2}) obtained when $y \ne y_0$ has
fundamental solutions given by
\begin{equation}
\varphi_1(\lambda,y) \equiv y \, F(a, \, b\,; \, c\,; \, y)
\ ,
\label{eq3.5}
\end{equation}
\begin{equation}
\varphi^*_1(\lambda,y) \equiv y^{-1/4} \, F(a-5/4, \, b-5/4\,;
\, 2-c\,; \, y)
\ ,
\label{eq3.6}
\end{equation}
where $F(a,b\,;c\,;z)$ denotes the hypergeometric function,$^{\ref{ref8}}$
and the parameters $a$, $b$, and $c$ are defined by
\begin{equation}
a \equiv {9 - \sqrt{17 + 16 \, \lambda} \over 8}\ , \ \ \ \ \ 
b \equiv {9 + \sqrt{17 + 16 \, \lambda} \over 8}\ , \ \ \ \ \ 
c \equiv {9 \over 4}
\ ,
\label{eq3.7}
\end{equation}
and therefore $a + b = c$.

\bigskip

\section*{\bf 3.1 Asymptotic Analysis}

\bigskip

The source photons injected into the flow are unable to diffuse very far
upstream due to the high speed of the inflowing electrons. Most of the
photons escape through the walls of the column within a few scattering
lengths of the source, and therefore we conclude that the function $g$
must {\it vanish} in the upstream limit, $y \to 0$. Asymptotic analysis
indicates that the function $\varphi_1(\lambda,y) \to 0$ in the limit $y
\to 0$ as required, but $\varphi^*_1(\lambda,y)$ diverges and therefore
it cannot be utilized in the upstream region ($y \le y_0$). Hence $g$
must be given by $\varphi_1$ for $y \le y_0$. Conversely, in the
downstream limit, the gas settles onto the surface of the star and
therefore $g$ should approach a constant as $y \to 1$. These conditions
are satisfied if $\lambda$ is equal to one of the eigenvalues,
$\lambda_n$, which are associated with the spatial eigenfunctions,
$g_n(y)$, defined by
\begin{equation}
g_n(y) \equiv g(\lambda_n,y)
\ .
\label{eq3.8}
\end{equation}
In order to obtain a complete understanding of the global behavior of
the eigenfunctions, we must also consider the asymptotic behaviors of
the two functions $\varphi_1$ and $\varphi_1^*$ in the downstream
region, which are discussed below.

The hypergeometric functions appearing in (\ref{eq3.5}) and
(\ref{eq3.6}) can be evaluated at $y=1$ using Eq.~(15.1.20) from
Abramowitz \& Stegun,$^{\ref{ref8}}$ which gives for general values of
$a$, $b$, and $c$
\begin{equation}
F(a\,,\,b\,;\,c\,;1) = {\Gamma(c) \, \Gamma(c-a-b) \over \Gamma(c-a)
\, \Gamma(c-b)}
\ .
\label{eq3.9}
\end{equation}
However, for the values of $a$, $b$, and $c$ in (\ref{eq3.5}) and
(\ref{eq3.6}), we find that [see Eq.~(\ref{eq3.7})]
\begin{equation}
c - a - b = 0
\ ,
\label{eq3.10}
\end{equation}
and therefore the hypergeometric functions $F(a\,, \, b\,; \, c\,; \,
y)$ and $F(a-5/4\,, \, b-5/4\,;\, 2-c\,; \, y)$ each {\it diverge} in
the downstream limit $y \to 1$. Since the eigenfunction $g_n$ should
approach a constant as $y \to 1$ based on physical considerations, we
conclude that in the downstream region ($y \ge y_0$), $g_n$ must be
represented by a suitable linear combination of $\varphi_1$ and
$\varphi^*_1$ that remains {\it finite} as $y \to 1$. In order to make
further progress, we need to employ Eq.~(15.3.10) from Abramowitz \&
Stegun,$^{\ref{ref8}}$ which yields for general $a$, $b$, and $y$
\begin{eqnarray}
F(a\,, b\,; a+b\,; y) = {\Gamma(a+b) \over \Gamma(a) \, \Gamma(b)}
\ \sum_{n=0}^\infty \ {(a)_n \, (b)_n \over (n!)^2} \bigg[2 \Psi(n+1)
- \Psi(a+n)
\nonumber
\\
- \Psi(b+n) - \ln(1-y)\bigg] (1-y)^n
\ ,
\label{eq3.11}
\end{eqnarray}
where
\begin{equation}
\Psi(z) \equiv {1 \over \Gamma(z) } \, {d \Gamma(z) \over dz}
\ .
\label{eq3.12}
\end{equation}
Asymptotic analysis of this expression reveals that in the limit $y \to
1$, the logarithmic divergences of the two functions $\varphi_1$ and
$\varphi^*_1$ can be balanced by creating the new function
\begin{equation}
\varphi_2(\lambda,y) \equiv {\Gamma(b) \over \Gamma(c) \, \Gamma(1-b)}
\ \varphi_1(\lambda,y)
- {\Gamma(1-a) \over \Gamma(2-c) \, \Gamma(a)}
\ \varphi^*_1(\lambda,y)
\ ,
\label{eq3.13}
\end{equation}
which remains finite as $y \to 1$. Hence $\varphi_2$ represents the
fundamental solution for $g_n$ in the region downstream from the source.
We can use the asymptotic behaviors of $\varphi_1$ and $\varphi^*_1$ to
show that
\begin{equation}
\lim_{y \to 1} \ \varphi_2(\lambda,y) = {\pi \, [\cot(\pi \, a)
+ \cot(\pi \, b)] \over \Gamma(a) \, \Gamma(1-b)}
\ .
\label{eq3.14}
\end{equation}

Since the solutions $\varphi_1$ and $\varphi_2$ are applicable
in the upstream and downstream regions, respectively, the global
expression for the eigenfunction $g_n$ is therefore given by
\begin{equation}
g_n(y) =
\begin{cases}
\varphi_1(\lambda_n,y) \ , & y \le y_0 \ , \cr
%
%
B_n \, \varphi_2(\lambda_n,y) \ , & y \ge y_0 \ , \cr
\end{cases}
\label{eq3.15}
\end{equation}
where the constant $B_n$ is evaluated using the continuity condition
[Eq.~(\ref{eq3.3})], which yields
\begin{equation}
B_n = {\varphi_1(\lambda_n,y_0) \over \varphi_2(\lambda_n,y_0)}
\ .
\label{eq3.16}
\end{equation}
It follows from (\ref{eq3.14}), (\ref{eq3.15}), and (\ref{eq3.16}) that
the downstream value of $g_n$ is given by
\begin{equation}
\lim_{y \to 1} \ g_n(y)
= {\pi \, [\cot(\pi \, a)
+ \cot(\pi \, b)] \over \Gamma(a) \, \Gamma(1-b)}
\, {\varphi_1(\lambda_n,y_0) \over \varphi_2(\lambda_n,y_0)}
\ .
\label{eq3.17}
\end{equation}
Conversely, in the upstream region, $\varphi_1 \to y$, and therefore
we have the asymptotic behavior
\begin{equation}
\lim_{y \to 0} \ {g_n(y) \over y} = 1
\ .
\label{eq3.18}
\end{equation}

\bigskip

\section*{\bf 3.2 Eigenvalue Equation}

\bigskip

We can combine (\ref{eq3.4}), (\ref{eq3.15}), and (\ref{eq3.16}) to show
that the eigenvalues $\lambda_n$ satisfy the equation
\begin{equation}
W(\lambda_n,y_0) - {3 \, \beta \, \varphi_1(\lambda_n,y_0) \,
\varphi_2(\lambda_n,y_0) \over 4 \, y_0} = 0
\ ,
\label{eq3.19}
\end{equation}
where the Wronskian of the two functions $\varphi_1$ and $\varphi_2$
is defined for general values of $\lambda$ and $y$ by
\begin{equation}
W(\lambda,y) \equiv
\varphi_1 \, {d \varphi_2 \over d y}
- \varphi_2 \, {d \varphi_1 \over d y}
\ .
\label{eq3.20}
\end{equation}
Further progress can be made by deriving an analytical expression for
the Wronskian. We begin by writing the differential
equation~(\ref{eq3.2}) governing the two functions $\varphi_1$ and
$\varphi_2$ in the self-adjoint form
\begin{equation}
{d \over dy}\left[y^{1/4} \, (1-y) \, {d\varphi \over dy}\right]
+ {\lambda \over 4 \, y^{3/4}} \, \varphi - T \, \varphi = 0
\ ,
\label{eq3.21}
\end{equation}
where
\begin{equation}
T \equiv {1-y \over 4 \, y^{7/4}} + {3 \, \beta \, v_0 \, \delta(y-y_0)
\over 7 \, v_c \, y^{3/4}}
\ .
\label{eq3.22}
\end{equation}
By applying (\ref{eq3.21}) to the function $\varphi_2$ and multiplying
the result by $\varphi_1$, and then subtracting from this the same
expression with $\varphi_1$ and $\varphi_2$ interchanged, we obtain
\begin{equation}
\varphi_1 \, {d \over dy}\left[y^{1/4} \, (1-y) \,
{d\varphi_2 \over dy}\right]
- \varphi_2 \, {d \over dy}\left[y^{1/4} \, (1-y) \,
{d\varphi_1 \over dy}\right] = 0
\ ,
\label{eq3.23}
\end{equation}
which can be rewritten as
\begin{equation}
y^{1/4} \, (1-y) \, {dW\over dy} + W \, {d\over dy}
\left[y^{1/4} \, (1-y)\right] = 0
\ ,
\label{eq3.24}
\end{equation}
where we have made use of the result
\begin{equation}
{dW \over dy} = \varphi_1 \, {d^2 \varphi_2 \over dy^2}
- \varphi_2 \, {d^2 \varphi_1 \over dy^2}
\ .
\label{eq3.25}
\end{equation}

Equation~(\ref{eq3.24}) can rearranged in the form
\begin{equation}
{d \ln W \over dy} = - {d \over dy} \ln\left[y^{1/4} \,
(1-y)\right]
\ ,
\label{eq3.26}
\end{equation}
which can be integrated to obtain the exact solution
\begin{equation}
W(\lambda,y) = {D(\lambda) \over y^{1/4} \, (1-y)}
\ ,
\label{eq3.27}
\end{equation}
where $D(\lambda)$ is an integration constant that depends on $\lambda$
but not on $y$. The exact dependence of $D$ on $\lambda$ can be derived
by analyzing the behaviors of the functions $\varphi_1$ and $\varphi_2$
in the limit $y \to 0$. For small values of $y$, we have the asymptotic
expressions$^{\ref{ref8}}$
\begin{eqnarray}
\varphi_1 &\to& y \ , \phantom{SPAAAAAAAAAACE} y \to 0 \ ,
\nonumber
\\
\varphi_2 &\to& - \, {\Gamma(1-a) \over \Gamma(a) \,
\Gamma(2-c)} \ y^{-1/4} \ , \ \ \ \ \ y \to 0 \ .
\label{eq3.28}
\end{eqnarray}
Combining (\ref{eq3.20}) and (\ref{eq3.28}), we find that
asymptotically,
\begin{equation}
W \to {5 \over 4} \, {\Gamma(1-a) \over \Gamma(a) \,
\Gamma(2-c)} \ y^{-1/4} \ , \ \ \ \ \ \ \ y \to 0
\ .
\label{eq3.29}
\end{equation}
Comparing this result with (\ref{eq3.27}), we conclude that
\begin{equation}
D(\lambda) = {5 \over 4} \, {\Gamma(1-a) \over \Gamma(a) \,
\Gamma(2-c)}
\ ,
\label{eq3.30}
\end{equation}
and therefore the exact solution for the Wronskian for general values
of $\lambda$ and $y$ is given by
\begin{equation}
W(\lambda,y) = {5 \over 4} \, {\Gamma(1-a) \over \Gamma(a) \,
\Gamma(2-c)} \ {y^{-1/4} \over 1-y}
\ .
\label{eq3.31}
\end{equation}
Substituting for $W$ in (\ref{eq3.19}) using (\ref{eq3.31}), we can
rewrite the eigenvalue equation in the equivalent form
\begin{equation}
{5 \over 3} \, {\Gamma(1-a) \over \Gamma(a) \, \Gamma(2-c)}
\, {y_0^{3/4} \over 1 - y_0} = \beta \, \varphi_1(\lambda_n,y_0)
\, \varphi_2(\lambda_n,y_0)
\ ,
\label{eq3.32}
\end{equation}
where $a$ and $b$ are functions of $\lambda_n$ by virtue of
(\ref{eq3.7}), and $c=9/4$. The roots of this expression are the
eigenvalues $\lambda_n$, and the associated eigenfunctions are evaluated
using (\ref{eq3.15}). The first eigenvalue, $\lambda_0$, is especially
important because it determines the power-law shape of the high-energy
portion of the Green's function [see Eq.~(\ref{eq3.1})].

In Figure~1 we plot the first eigenvalue $\lambda_0$ as a function of
the dimensionless parameters $\beta$ and $y_0$. Note that $\lambda_0$ is
a double-valued function of $y_0$ for fixed $\beta$, which is a
consequence of the imposed velocity profile [Eq.~(\ref{eq2.5})].
Physically, this behavior reflects the fact that it is always possible
to achieve a desired amount of compression (first-order Fermi
energization) by placing the source in a specific location in either the
upstream or downstream regions of the flow. We also observe that if we
increase the absorption parameter $\beta$ while holding $y_0$ fixed,
then $\lambda_0$ increases monotonically, and therefore the high-energy
spectrum becomes progressively steeper. This behavior is expected
physically because as the absorption parameter is increased, the
injected photons spend less time on average being energized by
collisions with electrons before either escaping from the column or
being absorbed at the source location. The decreased amount of
energization naturally leads to a steepening of the radiation spectrum.
When $\beta = 0$, no absorption occurs, and the index $\lambda_0$
achieves it minimum (limiting) value of 4. This limit is, however,
unphysical since it yields a divergent result for the total photon
energy density $\ugreen$ according to (\ref{eq2.2}). Nonetheless, the
case with $\beta=0$ is interesting from a mathematical viewpoint, and
for that reason it is further discussed in section~V.

\begin{figure}
\includegraphics[width=100mm]{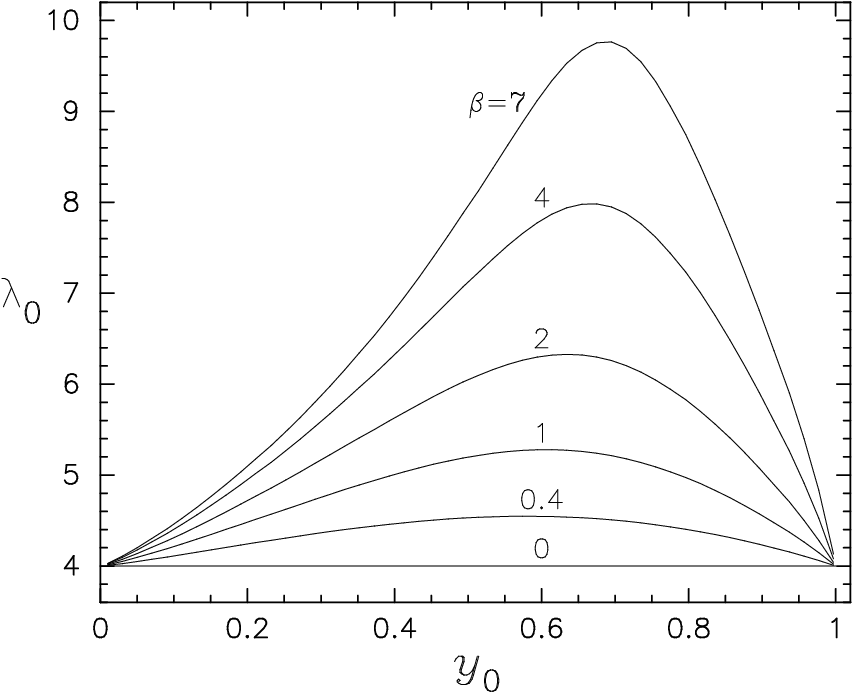}
\caption{First eigenvalue $\lambda_0$ of the Green's function expansion
plotted as a function of the source location $y_0$ for the indicated
values of the absorption parameter $\beta $. Note the steepening of the
radiation spectrum that occurs when $\beta$ is increased for a fixed
value of $y_0$, which reflects the decreasing residence time for the
photons in the plasma (see the discussion in the text).}
\end{figure}

\bigskip

\section*{\bf 3.3 Orthogonality of the Eigenfunctions}

\bigskip

We shall next demonstrate that the eigenfunctions $g_n$(y) form an
orthogonal set, which is an extremely useful property. This is a
standard Sturm-Liouville problem and therefore we follow the usual
procedure. Let us suppose that $g_n(y)$ and $g_m(y)$ are two
eigenfunctions corresponding to the distinct eigenvalues $\lambda_n$ and
$\lambda_m$, respectively. The functions $g_n$ and $g_m$ each satisfy
the differential equation~(\ref{eq3.2}), and therefore we can utilize
the self-adoint form to write [cf. Eq.~(\ref{eq3.21})]
\begin{equation}
g_m \left\{
{d \over dy}\left[y^{1/4} \, (1-y) \, {d g_n \over dy}\right]
+ {\lambda_n \over 4 \, y^{3/4}} \, g_n - T \, g_n
\right\} = 0
\ ,
\label{eq3.33}
\end{equation}
and
\begin{equation}
g_n \left\{
{d \over dy}\left[y^{1/4} \, (1-y) \, {d g_m \over dy}\right]
+ {\lambda_m \over 4 \, y^{3/4}} \, g_m - T \, g_m
\right\} = 0
\ ,
\label{eq3.34}
\end{equation}
where $T$ is given by (\ref{eq3.22}). Subtracting the second equation
from the first yields, after integrating by parts with respect to $y$
from $y=0$ to $y=1$,
\begin{equation}
(\lambda_n - \lambda_m) \int_0^1 y^{-3/4} \, g_n(y) \, g_m(y) \, dy
= 4 \, y^{1/4} \, (1-y) \left[
g_n \, {d g_m\over dy} - g_m \, {d g_n \over dy}
\right]\Bigg|_0^1
\ .
\label{eq3.35}
\end{equation}
Based on the asymptotic behaviors of the eigenfunctions $g_n$ and $g_m$
given by (\ref{eq3.17}) and (\ref{eq3.18}), we find that the right-hand
side of (\ref{eq3.35}) vanishes exactly, and therefore we obtain
\begin{equation}
(\lambda_n - \lambda_m) \int_0^1 y^{-3/4} \, g_n(y) \, g_m(y) \, dy
= 0
\ ,
\label{eq3.36}
\end{equation}
which establishes the orthogonality of the eigenfunctions. The set of
eigenfunctions is also complete according to the Sturm-Liouville
theorem. Since the eigenfunctions are orthogonal, the Green's function
can be expressed as the infinite series
\begin{equation}
\green(y_0,y,\epsilon_0,\epsilon) = \sum_{n=0}^\infty \ A_n
\left(\epsilon \over\epsilon_0\right)^{-\lambda_n} g_n(y)
\ ,
\label{eq3.37}
\end{equation}
for $\epsilon \ge \epsilon_0$, where the expansion coefficients $A_n$
are computed by employing the orthogonality of the eigenfunctions along
with the condition
\begin{equation}
\green(y_0,y,\epsilon_0,\epsilon)\bigg|_{\epsilon=\epsilon_0}
= {12 \, \dot N_0 \over 7 \, \pi \, r_0^2 \, \epsilon_0^3 \, v_c}
\ \delta(y - y_0)
\ ,
\label{eq3.38}
\end{equation}
which is obtained by integrating the transport equation~(\ref{eq2.11})
with respect to $\epsilon$ in a small range surrounding the injection
energy $\epsilon_0$. The result obtained for the $n$th expansion
coefficient is
\begin{equation}
A_n = {12 \, \dot N_0 \, y_0^{-3/4} \, g_n(y_0) \over
7 \, \pi \, r_0^2 \, \epsilon_0^3 \, v_c \, \mathfrak C_n}
\ ,
\label{eq3.39}
\end{equation}
where the quadratic normalization integrals, $\mathfrak C_n$, are
defined by
\begin{equation}
\mathfrak C_n \equiv \int_0^1 y^{-3/4} \, g_n^2(y) \, dy
\ .
\label{eq3.40}
\end{equation}
As an alternative to numerical integration, in section~3.4 we
derive a closed-form expression for evaluating the normalization
integrals based directly on the associated differential equation.

\bigskip

\section*{\bf 3.4 Quadratic Normalization Integrals}

\bigskip

The direct computation of the normalization integrals $\mathfrak C_n$
via numerical integration is costly and time consuming, and therefore it
is desirable to have an alternative procedure available for their
evaluation. In fact, it is possible to derive an analytical expression
for the normalization integrals based on manipulation of the fundamental
differential equation~(\ref{eq3.2}) governing the eigenfunctions
$g_n(y)$.

Let us suppose that $g(\lambda,y)$ is a general solution to
(\ref{eq3.2}) for an arbitrary value of $\lambda$ (i.e., not necessarily
an eigenvalue) with the asymptotic (upstream) behavior
\begin{equation}
g(\lambda,y) \to y \ , \ \ \ \ \ \ \ \ y \to 0
\ ,
\label{eq3.41}
\end{equation}
which is the same as the upstream behavior of the eigenfunction $g_n(y)$
[see Eq.~(\ref{eq3.18})]. We also stipulate that $g$ must be continuous
at $y=y_0$, and that it satisfies the derivative jump condition given by
(\ref{eq3.4}). After a bit of algebra, we find that the global solution
for $g$ consistent with these requirements can be expressed as
\begin{equation}
g(\lambda,y) =
\begin{cases}
\varphi_1(\lambda,y) \ , & y \le y_0 \ , \cr
%
%
(1+\hat a) \, \varphi_1(\lambda,y) + \hat b \, \varphi_2(\lambda,y)
\ , & y \ge y_0 \ , \cr
\end{cases}
\label{eq3.42}
\end{equation}
where the coefficients $\hat a$ and $\hat b$ are given by
\begin{equation}
\hat a = - \, {3 \, \beta \, \varphi_1(\lambda,y_0)
\, \varphi_2(\lambda,y_0)
\over 4 \, y_0 W(\lambda,y_0)} \ , \ \ \ \ \ \ \ 
\hat b = \ \ \ {3 \, \beta \, \varphi_1^2(\lambda,y_0) \over
4 \, y_0 W(\lambda,y_0)} \ ,
\label{eq3.43}
\end{equation}
and the Wronskian $W$ is evaluated using (\ref{eq3.31}).

Comparing the general solution for $g(\lambda,y)$ with the solution for
the eigenfunction $g_n(y)$ given by (\ref{eq3.15}), we note that
\begin{equation}
\lim_{\lambda \to \lambda_n} \hat a = -1 \ , \ \ \ \ \ \ 
\lim_{\lambda \to \lambda_n} \hat b = B_n
\ .
\label{eq3.44}
\end{equation}
We can now use the self-adoint form of (\ref{eq3.2}) to write [cf.
Eqs.~(\ref{eq3.33}) and (\ref{eq3.34})]
\begin{equation}
g_n \left\{
{\partial \over \partial y}\left[y^{1/4} \, (1-y) \,
{\partial g \over \partial y}\right]
+ {\lambda \over 4 \, y^{3/4}} \, g - T \, g
\right\} = 0
\ .
\label{eq3.45}
\end{equation}
and
\begin{equation}
g \left\{
{d \over dy}\left[y^{1/4} \, (1-y) \, {d g_n \over dy}\right]
+ {\lambda_n \over 4 \, y^{3/4}} \, g_n - T \, g_n
\right\} = 0
\ ,
\label{eq3.46}
\end{equation}
where $T$ is defined by (\ref{eq3.22}). Subtracting the second equation
from the first and integrating by parts from $y=0$ to $y=1$ yields
\begin{equation}
(\lambda - \lambda_n) \int_0^1 y^{-3/4} \, g(\lambda,y)
\, g_n(y) \, dy
= 4 \, y^{1/4} \, (1-y) \left[
g(\lambda,y) \, {d g_n\over dy} - g_n(y) \, {\partial g \over \partial y}
\right]\Bigg|_0^1
\ .
\label{eq3.47}
\end{equation}
Since $g \to y$ and $g_n \to y$ as $y \to 0$, we conclude that the
evaluation at the lower bound $y=0$ on the right-hand side yields zero,
and consequently in the limit $\lambda \to \lambda_n$ we obtain for the
quadratic normalization integral $\mathfrak C_n$ [see
Eq.~(\ref{eq3.40})]
\begin{equation}
\mathfrak C_n = \int_0^1 y^{-3/4} \, g_n^2(y) \, dy
= \lim_{\lambda \to \lambda_n}
{4 \, y^{1/4} \, (1-y) \left[
g(\lambda,y) \, (d g_n / dy) - g_n(y) \, (\partial g / \partial y)
\right] \over \lambda - \lambda_n}\Bigg|_{y=1}
\ .
\label{eq3.48}
\end{equation}

The numerator and denominator on the right-hand side of (\ref{eq3.48})
each vanish in the limit $\lambda \to \lambda_n$, and therefore
we can employ L'H\^opital's rule to show that (e.g., Becker$^{\ref{ref9}})$
\begin{equation}
\mathfrak C_n = \lim_{\lambda \to \lambda_n}
4 \, y^{1/4} \, (1-y) \left[
{\partial g \over \partial y} \, {d g_n \over dy} - g_n \,
{\partial^2 g \over \partial y \, \partial \lambda}
\right]\Bigg|_{y=1}
\ .
\label{eq3.49}
\end{equation}
Substituting the analytical forms for $g_n(y)$ and $g(\lambda,y)$ given
by (\ref{eq3.15}) and (\ref{eq3.42}), respectively, we find that
(\ref{eq3.49}) can be rewritten as
\begin{equation}
\mathfrak C_n = \lim_{y \to 1} \
4 \, y^{1/4} \, (1-y) \, B_n \left[
W(\lambda,y) \, {d \hat a \over d \lambda}
+ B_n \, {\partial\varphi_2 \over \partial \lambda} \,
{\partial\varphi_2 \over \partial y}
- B_n \, \varphi_2(\lambda,y) \, {\partial^2 \varphi_2 \over
\partial y \, \partial \lambda}\right]
\Bigg|_{\lambda=\lambda_n}
\ ,
\label{eq3.50}
\end{equation}
where we have also utilized (\ref{eq3.20}) and (\ref{eq3.44}). Based on
the asymptotic behavior of $\varphi_2$ [see~(\ref{eq3.14})], we conclude
that the final two terms on the right-hand side of (\ref{eq3.50})
contribute nothing in the limit $y \to 1$, and therefore our
expression for $\mathfrak C_n$ reduces to
\begin{equation}
\mathfrak C_n = \lim_{y \to 1} \
4 \, y^{1/4} \, (1-y) \, B_n \, W(\lambda,y)
{d \hat a \over d \lambda}
\Bigg|_{\lambda=\lambda_n}
\ .
\label{eq3.51}
\end{equation}

Since $y = 1$ is a singular point of the differential equation
(\ref{eq3.2}), it is convenient to employ the relation
[see Eq.~(\ref{eq3.27})]
\begin{equation}
W(\lambda,y) \ y^{1/4} \, (1-y)
= W(\lambda,y_0) \ y_0^{1/4} \, (1-y_0)
\ ,
\label{eq3.52}
\end{equation}
which allows us to transform the evaluation in (\ref{eq3.51}) from
$y=1$ to $y=y_0$ to obtain the equivalent result
\begin{equation}
\mathfrak C_n =
4 \, y_0^{1/4} \, (1-y_0) \, \hat a \, W(\lambda,y_0) \,
{\varphi_1(\lambda_n,y_0)
\over \varphi_2(\lambda_n,y_0)} \,
{d \ln \hat a \over d \lambda}
\Bigg|_{\lambda=\lambda_n}
\ ,
\label{eq3.53}
\end{equation}
where we have also substituted for $B_n$ using (\ref{eq3.16}). The
derivative on the right-hand side can be evaluated using (\ref{eq3.43}),
which yields
\begin{equation}
{d \ln \hat a \over d \lambda}
= {\partial\ln\varphi_1 \over \partial\lambda}
+ {\partial\ln\varphi_2 \over \partial\lambda}
- {\partial \ln W \over \partial \lambda}
\ ,
\label{eq3.54}
\end{equation}
where the derivative of the Wronskian is given by [see Eqs.~(\ref{eq3.7})
and (\ref{eq3.31})]
\begin{equation}
{\partial \ln W \over \partial \lambda}
= {\Psi(a) + \Psi(1-a) \over (17 + 16 \, \lambda)^{1/2}}
\ ,
\label{eq3.55}
\end{equation}
and
\begin{equation}
\Psi(z) \equiv {1 \over \Gamma(z)} \, {d \Gamma(z) \over dz}
\ .
\label{eq3.56}
\end{equation}
Combining (\ref{eq3.43}), (\ref{eq3.53}), (\ref{eq3.54}), and
(\ref{eq3.55}), we find that that the quadratic normalization integrals
can be evaluated using the closed-form expression
\begin{equation}
\mathfrak C_n = K(\lambda_n,y_0)
\ ,
\label{eq3.57}
\end{equation}
where
\begin{equation}
K(\lambda,y) \equiv
3 \, \beta \, y^{-3/4} (1-y) \, \varphi_1^2(\lambda,y)
\left[{\Psi(a) + \Psi(1-a) \over (17 + 16 \, \lambda)^{1/2}}
- {\partial\ln\varphi_1\over\partial\lambda}
- {\partial\ln\varphi_2\over\partial\lambda}
\right]
\ .
\label{eq3.58}
\end{equation}
This formula provides an extremely efficient alternative to numerical
integration for the computation of $\mathfrak C_n$.

\bigskip

\section*{\bf 3.5 Numerical Examples}

\bigskip

\begin{figure}
\includegraphics[width=100mm]{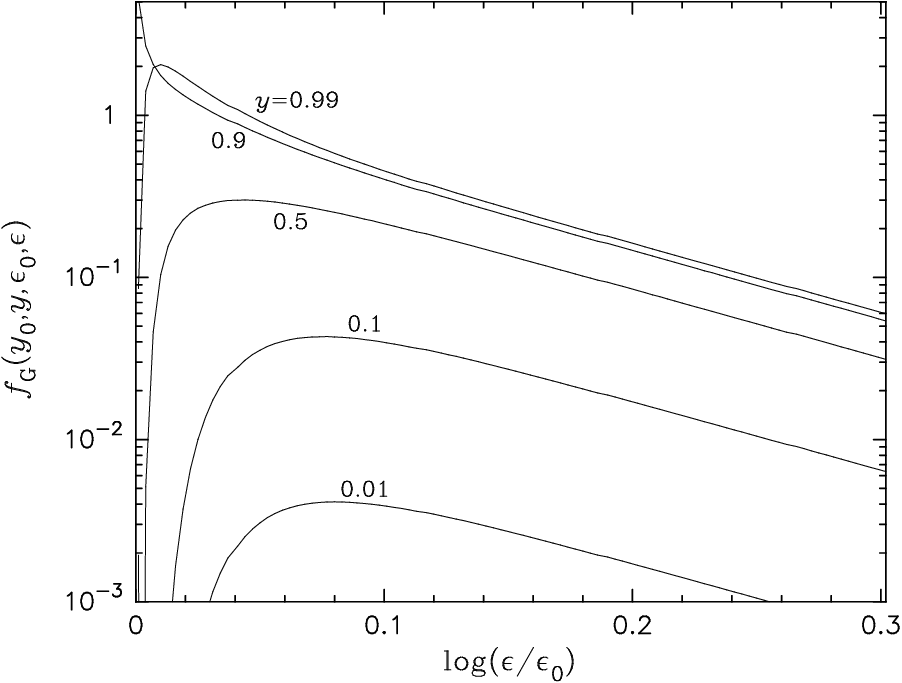}
\caption{Green's function $\green(y_0,y,\epsilon_0,\epsilon)$
[Eq.~(\ref{eq3.37})] plotted in units of $\dot N_0/(r_0^2 \epsilon_0^3
v_c)$ as a function of the photon energy ratio $\epsilon/\epsilon_0$ for
the indicated values of the spatial variable $y$. In this example we
have set the absorption constant $\beta = 0.4$ and the source location
parameter $y_0 = 0.9$, so that the source is located near the base of
the accretion column.}
\end{figure}

In this section we illustrate the computational method by examining the
dependence of the Green's function $\green(y_0,y,\epsilon_0,\epsilon)$
on the spatial location $y$ and the energy $\epsilon$. We remind the
reader that the solution for the Green's function represents the photon
spectrum inside the accretion column at the specified position and
energy, resulting from the injection of monochromatic photons with
energy $\epsilon_0$ from a source located at $y_0$. Hence analysis of
$\green$ allows us to explore the competing effects of Fermi
energization and diffusion as photons travel through the column. The
Green's function can be computed by combining (\ref{eq3.37}),
(\ref{eq3.39}), and (\ref{eq3.57}) once the eigenvalues $\lambda_n$ have
been determined using (\ref{eq3.32}). The eigenfunction expansion for
$\green$ converges fairly rapidly, and in general one obtains at least
five decimal digits of accuracy if the series in (\ref{eq3.37}) is
terminated after the first 20 terms.

The Green's function $\green(y_0,y,\epsilon_0,\epsilon)$ is plotted as a
function of the energy ratio $\epsilon/\epsilon_0$ and the location $y$
in Figure~2 for the parameter values $\beta = 0.4$ and $y_0 = 0.9$. In
this case the first eigenvalue is given by $\lambda_0 = 4.231$ (see
Fig.~1), which is equal to the high-energy slope of the Green's function
in the log-log plots in Fig.~2. The selected value of $y_0$ corresponds
to a source located near the bottom of the accretion column, just above
the stellar surface. At the source location, $y=y_0=0.9$, the energy
spectrum extends down to the injection energy, $\epsilon_0$. However, at
all other radii the spectrum displays a steep turnover above that energy
because all of the photons have experienced Fermi energization due to
collisions with the infalling electrons. The photons with energy
$\epsilon = \epsilon_0$ at the source location have been injected so
recently that they have not yet experienced significant energization.
Note that in the far upstream region (i.e., for small values of $y$),
the spectrum is greatly attenuated due to the inability of the photons
to diffuse upstream through the rapidly infalling plasma. In this
example, the average photon energy achieves its maximum value in the
upstream region because these are the photons that have resided in the
flow the longest and therefore experienced the most energy
amplification. However, due to the attenuation mentioned above, there
are not many of these photons.

In Fig.~3 we plot the Green's function $\green$ for the case with $\beta
= 4$ and $y_0 = 0.4$, which yields for the first eigenvalue $\lambda_0 =
6.325$. The source is now located in the upstream region and the
absorption is stronger, and consequently the behavior is somewhat
different from that displayed in Fig.~2. In particular, the photons
experience less overall compression in the flow and therefore the
spectrum is steeper at high energies, as evidenced by the increase in
the primary eigenvalue $\lambda_0$. This is mainly due to the larger
value of $\beta$, which causes the photons to spend less time on average
in the flow being energized by collisions with the electrons before they
escape from the column or are ``recycled'' by absorption. We also note
that in this case the average radiation energy displays its maximum value
in the downstream region. This is the reverse of the behavior displayed
in Fig.~2 because in the present situation, the source is located in the
upstream region and therefore the photons that diffuse further
upstream do not experience as much energization as those considered in
Fig.~2. The radiation distribution in the far upstream region is greatly
attenuated due to diffusion against the current of infalling electrons,
as in Fig.~2. The analytical results for the Green's function obtained
here provide the basis for the consideration of any source distribution
since the fundamental differential equation (\ref{eq2.1}) is linear.
This is further discussed in section~VI.

\begin{figure}
\includegraphics[width=100mm]{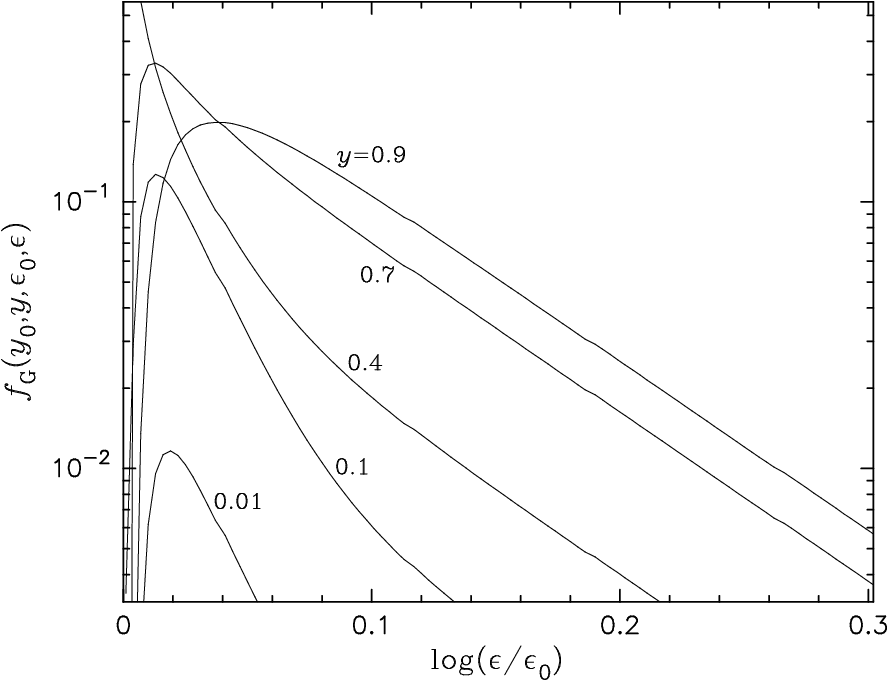}
\caption{Same as Fig.~2, except $\beta = 0.4$ and $y_0 = 0.9$. In this
case the source is located in the upstream region, and the average
photon energy achieves its maximum value in the downstream region.}
\end{figure}

\bigskip

\section*{\bf IV. HYPERGEOMETRIC SUMMATION FORMULA}

\bigskip

We can derive two interesting summation formulas for the hypergeometric
eigenfunctions by using the transport equation~(\ref{eq2.11}) to study
the behavior of the ``energy moments,'' $I_\ell$, defined by
\begin{eqnarray}
I_\ell(y) \equiv \int_{\epsilon_0}^\infty \epsilon^\ell \, \green
\, d\epsilon
\ .
\label{eq4.1}
\end{eqnarray}
The lower bound of $\epsilon_0$ is chosen because $\green = 0$ for
$\epsilon < \epsilon_0$ as explained in the discussion preceding
(\ref{eq3.1}). Note that according to (\ref{eq2.2}), the number and
energy densities are given by $\ngreen = I_2$ and $\ugreen = I_3$,
respectively. The differential equation satisfied by $I_\ell$ is
obtained by operating on (\ref{eq2.11}) with $\int \epsilon^\ell \,
d\epsilon$, which yields
\begin{equation}
y \, (1-y) \, {d^2 I_\ell \over d y^2}
+ \left({1 - 5 \, y \over 4}\right) {d I_\ell \over d y}
+ \left(\ell y + 2 y - 1 \over 4 y \right) I_\ell
= {3 \, \beta \, v_0 \, \delta(y-y_0) \, I_\ell \over 7 \, v_c}
- {3 \, \dot N_0 \, \epsilon_0^{\ell-2} \, \delta(y-y_0)
\over 7 \, \pi \, r_0^2 \, v_c}
\ .
\label{eq4.2}
\end{equation}
The energy moment $I_\ell$ must be continuous at $y=y_0$ in order to
avoid generating an infinite spatial diffusion flux there, and
consequently we have
\begin{equation}
\Delta\left[I_\ell(y)\right]\Bigg|_{y=y_0} = 0
\ .
\label{eq4.3}
\end{equation}
By integrating (\ref{eq4.2}) in a small region around $y=y_0$, we can
show that $I_\ell$ also satisfies the derivative jump condition
\begin{equation}
\Delta\left[{d I_\ell \over d y}\right]\Bigg|_{y=y_0}
= {3 \, \beta \, I_\ell(y_0)
\over 4 \, y_0}
- {3 \, \dot N_0 \, \epsilon_0^{\ell-2}
\over 7 \, \pi \, r_0^2 \, v_c \, y_0 \, (1-y_0)}
\ .
\label{eq4.4}
\end{equation}
where we have also utilized (\ref{eq2.12}).

The homogeneous version of (\ref{eq4.2}) obtained when $y \ne y_0$ is
equivalent to (\ref{eq3.2}) for $g$ if we replace $\lambda$ with
$\ell+1$. Since the energy moments $I_\ell$ must satisfy the same
upstream and downstream boundary conditions that apply to the separation
eigenfunctions $g_n$, we can therefore write the general solution for
$I_\ell$ as
\begin{equation}
I_\ell(y) =
\begin{cases}
C_\ell \, \varphi_1(\ell+1,y) \ , & y \le y_0 \ , \cr
%
%
D_\ell \, \varphi_2(\ell+1,y) \ , & y \ge y_0 \ , \cr
\end{cases}
\label{eq4.5}
\end{equation}
where the constants $C_\ell$ and $D_\ell$ are computed by satisfying the
continuity and derivative jump conditions given by (\ref{eq4.3})
and (\ref{eq4.4}). Upon substitution, we obtain after some algebra
\begin{equation}
C_\ell  = {12 \, \dot N_0 \, \epsilon_0^{\ell-2} \over 7 \pi
\, v_c \, r_0^2} {(1-y_0)^{-1} \, \varphi_2(\ell+1,y_0)
\over 3 \, \beta \, \varphi_1(\ell+1,y_0) \, \varphi_2(\ell+1,y_0)
- 4 \, y_0 W(\ell+1,y_0)}
\ ,
\label{eq4.6}
\end{equation}
\begin{equation}
D_\ell  = {12 \, \dot N_0 \, \epsilon_0^{\ell-2} \over 7 \pi
\, v_c \, r_0^2} {(1-y_0)^{-1} \, \varphi_1(\ell+1,y_0)
\over 3 \, \beta \, \varphi_1(\ell+1,y_0) \, \varphi_2(\ell+1,y_0)
- 4 \, y_0 W(\ell+1,y_0)}
\ ,
\label{eq4.7}
\end{equation}
where $W(\ell+1,y_0)$ is computed using [cf.~Eq.(\ref{eq3.31})]
\begin{equation}
W(\ell+1,y_0) = {5 \over 4} \, {\Gamma(1-a_\ell) \over \Gamma(a_\ell) \,
\Gamma(-1/4)} \ {y_0^{-1/4} \over 1-y_0}
\ ,
\label{eq4.8}
\end{equation}
and
\begin{equation}
a_\ell \equiv {9 - \sqrt{33 + 16 \, \ell} \over 8}
\ .
\label{eq4.9}
\end{equation}

The energy moments $I_\ell(y)$ can also be calculated by substituting
for the Green's function in the fundamental integral (\ref{eq4.1}) using
(\ref{eq3.37}). Reversing the order or summation and integration yields
\begin{equation}
I_\ell(y) = \epsilon_0^{\ell+1}
\sum_{n=0}^\infty \ A_n (\lambda_n - \ell -1)^{-1}
\, g_n(y)
\ ,
\label{eq4.10}
\end{equation}
where $g_n(y)$ and $A_n$ are given by (\ref{eq3.15}) and (\ref{eq3.39}),
respectively. Note that the expression for $g_n(y)$ can be rewritten as
\begin{equation}
g_n(y) = {\varphi_1(\lambda_n,\ymin) \, \varphi_2(\lambda_n,\ymax)
\over \varphi_2(\lambda_n,y_0)}
\ ,
\label{eq4.11}
\end{equation}
where
\begin{equation}
\ymin \equiv {\rm min}(y,y_0) \ , \ \ \ \ \ \ \ 
\ymax \equiv {\rm max}(y,y_0)
\ .
\label{eq4.12}
\end{equation}
Eliminating $I_\ell(y)$ between (\ref{eq4.5}) and (\ref{eq4.10}) and
making use of (\ref{eq3.39}), (\ref{eq4.6}), (\ref{eq4.7}), and
(\ref{eq4.11}), we find after some simplification that
\begin{equation}
\sum_{n=0}^\infty \ {\varphi_1(\lambda_n,y_0) \over
\varphi_2(\lambda_n,y_0)} \, {\varphi_1(\lambda_n,\ymin)
\, \varphi_2(\lambda_n,\ymax) \over (\lambda_n - \ell - 1)
\, \mathfrak C_n}
= {y_0^{3/4} \, (1-y_0)^{-1} \, \varphi_1(\ell+1,\ymin)
\, \varphi_2(\ell+1,\ymax) \over 3 \, \beta \, \varphi_1(\ell+1,y_0)
\, \varphi_2(\ell+1,y_0) - 4 \, y_0 W(\ell+1,y_0)}
\ ,
\label{eq4.13}
\end{equation}
where the eigenvalues $\lambda_n$ are computed using (\ref{eq3.32}).
Equation (\ref{eq4.13}) is a new hypergeometric summation formula that
has not appeared previously in the literature. This relation holds for
all real values of $\ell$.

\bigskip

\section*{\bf V. LINEAR AND BILINEAR GENERATING FUNCTIONS}

\bigskip

The case with $\beta=0$ is interesting from a mathematical point of view
because in this limit, the hypergeometric eigenfunctions reduce to
Jacobi polynomials. We can therefore combine various results from
sections~III and IV to obtain two new summation formulas (i.e., linear
and bilinear generating functions) for the Jacobi polynomials that have
not appeared previously in the literature. In the limit $\beta \to 0$,
the eigenvalue equation (\ref{eq3.19}) reduces to
\begin{equation}
W(\lambda_n,y_0)
= {5 \over 4} \, {\Gamma(1-a) \over \Gamma(a) \,
\Gamma(-1/4)} \ {y_0^{-1/4} \over 1-y_0} = 0
\ ,
\label{eq5.1}
\end{equation}
where we have also made use of (\ref{eq3.31}). Roots of this expression
occur where $|\Gamma(a)| \to \infty$, which corresponds to
\begin{equation}
a = -n \ , \ \ \ \ \ \ \ n = 0, 1, 2, \ldots
\label{eq5.2}
\end{equation}
In this situation, we can use (\ref{eq3.7}) to demonstrate that the
exact solution for the eigenvalues $\lambda_n$ is given by
\begin{equation}
\lambda_n = 4 \, n^2 + 9 \, n + 4
\ .
\label{eq5.3}
\end{equation}
Next we note that $a + b = 9/4$ in general according to (\ref{eq3.7}),
and therefore we find that
\begin{equation}
b = {9 \over 4} + n
\ .
\label{eq5.4}
\end{equation}
The corresponding expression for the fundamental upstream eigensolution,
$\varphi_1(\lambda_n,y)$, is given in this case by the polynomial [see
Eq.~(\ref{eq3.5})]
\begin{equation}
\varphi_1(\lambda_n,y) = y \, F\left(-n, \, {9 \over 4} + n\,;
\, {9 \over 4}\,; \, y \right)
\ ,
\label{eq5.5}
\end{equation}
and the fundamental eigensolution in the downstream region,
$\varphi_2(\lambda_n,y)$, likewise reduces to [see Eq.~(\ref{eq3.13})]
\begin{equation}
\varphi_2(\lambda_n,y) = {\Gamma(n+9/4) \over \Gamma(9/4) \,
\Gamma(-n-5/4)}
\ \varphi_1(\lambda,y)
\ .
\label{eq5.6}
\end{equation}
Hence the two eigensolutions $\varphi_1(\lambda_n,y)$ and
$\varphi_2(\lambda_n,y)$ are {\it linearly dependent functions} in this
case, which is expected since the Wronskian $W(\lambda_n,y_0)=0$
according to (\ref{eq5.1}). This in turn reflects the fact that there is
no derivative jump in the global separation eigenfunction $g_n(y)$ at $y
= y_0$ when $\beta=0$ [see Eq.~(\ref{eq3.4})].

Due to the linear dependence of $\varphi_1(\lambda_n,y)$ and
$\varphi_2(\lambda_n,y)$, equation~(\ref{eq4.11}) for the global
eigenfunction $g_n(y)$ now simplifies to
\begin{equation}
g_n(y) = \varphi_1(\lambda_n,y)
\ ,
\label{eq5.7}
\end{equation}
and therefore the summation formula presented in (\ref{eq4.13}) can be
rewritten in the $\beta = 0$ case as
\begin{equation}
\sum_{n=0}^\infty \ {\varphi_1(\lambda_n,y_0) \,
\varphi_1(\lambda_n,y) \over (\lambda_n - \ell - 1)
\, \mathfrak C_n}
= - \ {\varphi_1(\ell+1,\ymin)
\, \varphi_2(\ell+1,\ymax) \over 4 \, y_0^{1/4} (1-y_0) \, W(\ell+1,y_0)}
\ ,
\label{eq5.8}
\end{equation}
where $\ymin$ and $\ymax$ are defined by (\ref{eq4.12}) and
$W(\ell+1,y_0)$ is computed using (\ref{eq4.8}).

We are now in a position to derive an interesting summation formula for
products of Jacobi polynomials. Using Eq.~(15.4.6) from Abramowitz \&
Stegun,$^{\ref{ref8}}$ our expression for the eigensolution
$\varphi_1(\lambda_n,y)$ can be rewritten as
\begin{equation}
\varphi_1(\lambda_n,y) = {n! \over (9/4)_n} \ y \, P_n^{(5/4,\,0)}
(1-2y)
\ ,
\label{eq5.9}
\end{equation}
where
\begin{equation}
P_n^{(5/4,\,0)}(1-2y)
= {(9/4)_n \over n!} \, F\left(-n, \, {9 \over 4} + n\,;
\, {9 \over 4}\,; \, y \right)
\label{eq5.10}
\end{equation}
represents the Jacobi polynomial, and $(a)_n$ denotes the Pochhammer
symbol, defined by$^{\ref{ref8}}$
\begin{equation}
(a)_n \equiv {\Gamma(a+n) \over \Gamma(a)}
\ .
\label{eq5.11}
\end{equation}
In the present application, with $\beta=0$, we can combine (\ref{eq3.40}),
(\ref{eq5.7}), and (\ref{eq5.9}) to express the quadratic normalization
integrals, $\mathfrak C_n$, as
\begin{equation}
\mathfrak C_n
= \left[{n! \over (9/4)_n}\right]^2
\int_0^1 y^{5/4} \left[P_n^{(5/4,\,0)}(1-2y)
\right]^2 \, dy
\ ,
\label{eq5.12}
\end{equation}
which can be evaluated using Eq.~(7.391.1) from Gradshteyn and
Ryzhik$^{\ref{ref10}}$ to obtain
\begin{equation}
\mathfrak C_n
= \left[{n! \over (9/4)_n}\right]^2
\left(2 \, n + {9 \over 4}\right)^{-1}
\ .
\label{eq5.13}
\end{equation}
Equations (\ref{eq4.8}), (\ref{eq5.3}), (\ref{eq5.8}), (\ref{eq5.9}),
and (\ref{eq5.13}) can be combined to derive a new {\it bilinear
generating function} for the Jacobi polynomials, which can be written as
\begin{equation}
\sum_{n=0}^\infty \ (9 + 8 n)
\, {P_n^{(5/4,\,0)}(1-2y_0) \, P_n^{(5/4,\,0)}(1-2y)
\over 4 n^2 + 9 n + 3 - \ell}
= {16 \over 5} \, {\Gamma(3/4) \, \Gamma(a_\ell) 
\over \Gamma(1-a_\ell)} \, {\varphi_1(\ell+1,\ymin)
\, \varphi_2(\ell+1,\ymax) \over y \, y_0}
\ ,
\label{eq5.14}
\end{equation}
where $a_\ell$ is defined by (\ref{eq4.9}). Note that the functions
$\varphi_1(\ell+1,\ymin)$ and $\varphi_2(\ell+1,\ymax)$ appearing on the
right-hand side of (\ref{eq5.14}) are {\it not} eigenfunctions since in
general the quantity $\ell + 1$ is not equal to one of the eigenvalues
$\lambda_n$.

An interesting special case occurs in the limit $y_0 \to 0$. Making
use of the relation [see Eq.~(\ref{eq5.10})]
\begin{equation}
P_n^{(5/4,\,0)}(1)
= {(9/4)_n \over n!}
\label{eq5.15}
\ ,
\end{equation}
and the identity
\begin{equation}
\Gamma\left({3 \over 4}\right) \, \Gamma\left({9 \over 4}\right)
= {5 \over 16} \, \pi \, 2^{1/2}
\ ,
\label{eq5.16}
\end{equation}
we now find that (\ref{eq5.14}) reduces to the {\it linear generating
function}
\begin{equation}
\sum_{n=0}^\infty \ {(9 + 8 n) \, \Gamma(n + 9/4) \over
(4 n^2 + 9 n + 3 - \ell) \, n!}
\ P_n^{(5/4,\,0)}(1-2y)
= {\pi \, 2^{1/2} \, \Gamma(a_\ell) 
\over \Gamma(1-a_\ell)} \, {\varphi_2(\ell+1,y) \over y}
\ ,
\label{eq5.17}
\end{equation}
which is valid for all real values of $\ell$. Equations (\ref{eq5.14})
and (\ref{eq5.17}) are new results that are useful for the evaluation of
infinite sums containing either products of Jacobi polynomials or single
Jacobi polynomials, respectively.

\section*{\bf VI. CONCLUSION}

In this article we have employed methods of classical analysis to obtain
the exact solution for the Green's function describing the Fermi
energization of photons scattered by infalling electrons in a pulsar
accretion column. This process is of central importance in the
development of theoretical models for the production of the X-ray
spectra observed from these objects, which are among the brightest
sources in the Milky Way galaxy. As demonstrated in Fig.~1 and equation
(\ref{eq3.37}), the Green's function is characterized by a power-law
shape at high photon energies, which is typical for a Fermi process. In
this scenario, photons gain their energy by diffusing back and forth
across the shock many times. The probability of multiple shock crossings
decreases exponentially with the number of crossings, and the mean
energy of the photons increases exponentially with the number of
crossings. This combination of factors naturally gives rise to a
power-law energy distribution.$^{\ref{ref11}}$ Hence shock energization
in the pulsar accretion column provides a natural explanation for the
spectrum of the high-energy radiation produced by X-ray pulsars.
Specific examples of the Green's function are plotted in Figs.~2 and 3.

Due to the linearity of the transport equation (\ref{eq2.1}), we can
employ the Green's function to calculate the radiation spectrum inside
the accretion column resulting from an arbitrary source spectrum
using the convolution$^{\ref{ref12}}$
\begin{equation}
f(y_0,y,\epsilon) = \int_0^\infty
j(\epsilon_0) \,
{\green(y_0,y,\epsilon_0,\epsilon) \over \dot N_0}
\, d \epsilon_0
\ ,
\label{eq6.1}
\end{equation}
where $j(\epsilon_0) \, d\epsilon_0$ represents the number of photons
injected into the accretion column between at location $y_0$ with energy
between $\epsilon_0$ and $\epsilon_0 + d \epsilon_0$. The source
distribution of greatest astrophysical interest is the ``thermal mound''
source located near the base of the accretion column, where the gas has
decelerated almost to rest and is therefore extremely dense. This hot
plasma is in full thermodynamic equilibrium, and consequently it
radiates a blackbody spectrum.$^{\ref{ref1}}$ The absorption parameter
$\beta$ has been included in the transport equation (\ref{eq2.1}) in
order to account for the fact that a blackbody acts as both a source and
a sink of radiation.$^{\ref{ref4}}$ The fundamental results for the
Green's function obtained in the present article will be used to study
the reprocessing of the blackbody radiation emitted from the thermal
mound in a subsequent paper.

In addition to the analytical results for the Green's function, we have
also obtained an interesting formula for the evaluation of an infinite
series involving products of the orthogonal hypergeometric
eigenfunctions [see Eq.~(\ref{eq4.13})]. This derivation was based on
the simultaneous calculation of the energy moments $I_\ell(y)$ using
either an expression based on term-by-term integration of the Green's
function expansion (\ref{eq3.37}), or an independent solution developed
via direct integration of the fundamental transport equation
(\ref{eq2.1}). In the special case $\beta \to 0$, which corresponds
physically to the neglect of absorption at the source location, our
general formula for the hypergemetric summation reduces to a bilinear
generating function for the Jacobi polynomials given by (\ref{eq5.14}).
This relation in turn simplifies to yield a linear generating function
for the Jacobi polynomials in the limit $y_0 \to 0$, which corresponds
physically to a source located in the far upstream region [see
Eq.~(\ref{eq5.17})].

The results derived is this article for the linear and bilinear
generating functions of Jacobi polynomials are related to various
similar expressions obtained previously by Chen and
Srivastava,$^{\ref{ref13},\,\ref{ref14}}$ Srivastava,$^{\ref{ref15}}$
Rangarajan,$^{\ref{ref16}}$ and Pittaluga, Sacripante, and
Srivastava.$^{\ref{ref17}}$ However, our results are not identical to
any of their formulas and therefore they represent an interesting new
family of relations. Although the linear and bilinear generating
functions developed here relate specifically to the properties of the
polynomials $P_n^{(5/4,\,0)}(1-2y)$, we expect that some level of
generalization may be possible. We plan to pursue this question in
future work.

\section*{REFERENCES}

\newcounter{refcount}
\begin{list}{$^{\arabic{refcount}}$\!\!}
{\usecounter{refcount}
\setlength{\leftmargin}{0.09in}
\setlength{\rightmargin}{\leftmargin}}

\item{\label{ref1}}
K. Davidson, ``Accretion at a magnetic pole of a neutron
star,'' Nat. Phys. Sci. {\bf 246}, 1--4 (1973).

\smallskip\noindent
\item{\label{ref2}}
P. A. Becker, ``First-order Fermi acceleration in spherically
symmetric flows: solutions including quadratic losses,'' Astrophys.
J. {\bf 397}, 88--116 (1992).

\smallskip\noindent
\item{\label{ref3}}
P. A. Becker and M. T. Wolff, ``Spectral formation in X-ray pulsar
accretion columns,'' Astrophys. J. Lett. {\bf 621}, L45--L48 (2005).

\smallskip\noindent
\item{\label{ref4}}
G. B. Rybicki and A. P. Lightman, {\it Radiative Processes
in Astrophysics} (Wiley, NY, 1979).

\smallskip\noindent
\item{\label{ref5}}
P. A. Becker, ``Dynamical structure of radiation-dominated pulsar
accretion shocks,'' Astrophys. J. {\bf 498}, 790--801 (1998).

\smallskip\noindent
\item{\label{ref6}}
R. D. Blandford and D. G. Payne, ``Compton scattering in a converging fluid
flow -- II. Radiation-dominated shock,'' Monthly Not. Royal Astron. Soc.
{\bf 194}, 1041--1055 (1981).

\smallskip\noindent
\item{\label{ref7}}
M. M. Basko and R. A. Sunyaev, ``The limiting luminosity
of accreting neutron stars with magnetic fields,'' Monthly
Not. Royal Astr. Soc. {\bf 175}, 395--417 (1976).

\smallskip\noindent
\item{\label{ref8}}
M. Abramowitz and I. A. Stegun, {\it Handbook of Mathematical
Functions} (Dover, New York, 1970).

\smallskip\noindent
\item{\label{ref9}}
P. A. Becker, ``Normalization integrals of orthogonal Heun
functions,'' J. Math. Phys. {\bf 38}, 3692--3699 (1997).

\smallskip\noindent
\item{\label{ref10}}
I. S. Gradshteyn and I. M. Ryzhik, {\it Table of Integrals,
Series, and Products} (Academic Press, London, 1980).

\smallskip\noindent
\item{\label{ref11}}
E. Fermi, ``Galactic magnetic fields and the origin of
cosmic radiation,'' Astrophys. J. {\bf 119}, 1--6 (1954).

\smallskip\noindent
\item{\label{ref12}}
P. A. Becker, ``Exact Solution for the Green's Function Describing
Time-Dependent Thermal Comptonization,'' Monthly Not. Royal Astr. Soc.
{\bf 343}, 215--240 (2003).

\smallskip\noindent
\item{\label{ref13}}
M.-P. Chen and H. M. Srivastava, ``Some extensions of Bateman's
product formulas for the Jacobi polynomials,'' J. Appl. Math. Stochast.
Anal. {\bf 8}, 423--428 (1995).

\smallskip\noindent
\item{\label{ref14}}
M.-P. Chen and H. M. Srivastava, ``Orthogonality relations and
generating functions for jacobi polynomials and related hypergeometric
functions,'' Appl. Math. and Comp. {\bf 68}, 153--188 (1995).

\smallskip\noindent
\item{\label{ref15}}
H. M. Srivastava, ``Some binlinear generating functions,''
Proc. U.S. Nat. Acad. Sci. {\bf 64}, 462-465 (1969).

\smallskip\noindent
\item{\label{ref16}}
S. K. Rangarajan, ``Bilinear generating functions for the Jacobi
polynomials. I.,'' Bull. Acad. Polon. Sci. S\'er. Sci. Math. Astronom. Phys.
{\bf 13}, 101--103 (1965).

\smallskip\noindent
\item{\label{ref17}}
G. Pittaluga, L. Sacripante, and H. M. Srivastava, ``Some families
of generating functions for the Jacobi and related othogonal polynomials,''
J. Math. Anal. App. {\bf 238}, 385--417 (1999).

\end{list}

\end{document}